\documentclass[journal=jacsat,manuscript=article]{achemso}

\usepackage[version=3]{mhchem} 
\usepackage{color}
\usepackage{todonotes}
\usepackage[normalem]{ulem}

\author{Nils le Coutre}
\affiliation{Institute of Physics, Universität Rostock, Rostock}
\altaffiliation{These authors contributed equally to this work}\author{Tolibjon Abdurakhmonov}
\affiliation{Institute of Physics, Universität Rostock, Rostock}
\altaffiliation{These authors contributed equally to this work}
\author{Paul Weinbrenner}
\affiliation{Institute of Physics, Universität Rostock, Rostock}
\author{Kenji Watanabe}
\affiliation{Research Center for Electronic and Optical Materials, National Institute for Materials Science, Tsukuba,
Japan}
\author{Takashi Taniguchi}
\affiliation{Research Center for Materials Nanoarchitectonics, National Institute for Materials Science, Tsukuba, Japan}
\author{Tobias Korn}
\affiliation{Institute of Physics, Universität Rostock, Rostock}
\author{Franziska Fennel}
\affiliation{Institute of Physics, Universität Rostock, Rostock}
\author{Oliver Kühn}
\affiliation{Institute of Physics, Universität Rostock, Rostock}
\alsoaffiliation{University of Rostock, Department of Life, Light and Matter, 18059 Rostock, Germany}
\author{Friedemann Reinhard}
\affiliation{Institute of Physics, Universität Rostock, Rostock}
\alsoaffiliation{University of Rostock, Department of Life, Light and Matter, 18059 Rostock, Germany}
\alsoaffiliation{Munich Center for Quantum Science and Technology (MCQST), 80799 Munich, Germany}
\email{friedemann.reinhard@uni-rostock.de}

\title{Growth of few-layer molecular crystals of PTCDI on hexagonal boron nitride by microspacing air-gap sublimation.}

\begin{document}







\begin{abstract}

Extended two-dimensional (2D) crystals of dye molecules adsorbed on 2D material substrates like boron nitride have recently become a subject of intense study, with potential applications ranging from quantum technology to optoelectronics. The most established technique for the production of these films is physical vapor transport in vacuum. 
We demonstrate that few-layer crystalline films of the organic dye molecule PTCDI on boron nitride can be produced by microspacing in-air sublimation, a radically simplified technique, not requiring complicated vacuum systems. The resulting layers display clearly resolved 
atomic step terraces in atomic force microscopy, and a clear polarization anisotropy in their fluorescence, confirming molecular alignment and long-range order. 
Using density functional theory and classical molecular dynamics simulations, the canted motive is identified as the most likely building block for the morphology of a PTDCI monolayer on the hBN substrate. 
\end{abstract}
\section*{Keywords}
Layered growth, PTCDI, Fluorescence, Polarization, Physical Vapor Deposition, hBN, DFT
\section{Introduction}
When planar dye molecules are adsorbed on a surface, they often arrange in extended 2D molecular crystals, i.e. long-range-ordered periodic assemblies of molecules \cite{fraboni16}. These molecular films promise a wide range of applications, including organic thin film transistors and opto-electronic devices like solar cells, photodetectors and electroluminescence displays. 
The formation of long-range order is greatly assisted if adsorption takes place on a structured substrate like graphene or hexagonal boron nitride (hBN). In fact, for a typical perylene-like molecule, the hexagonal atomic structure is nearly commensurate with the molecular structure, enforcing alignment of the adsorbed molecules~\cite{auwarter19, feyter03, alkhamisi18}. Molecular crystals adsorbed on these substrates have been a workhorse platform for scanning tunneling microscopy~\cite{feyter03} and thin film electronics\cite{svatek20}. 
Intriguingly, organic molecules adsorbed on or below boron nitride are also conjectured to be the source of single photon emission and, possibly, optically readable electron spin qubits in boron nitride~\cite{neumann23, chejanovsky21, stern22}, suggesting that a better control and understanding of molecular adsorption could pave the way to controlled and scalable creation of optically readably spin qubits.\par 
One prototypical system for the study of molecular crystals is a class of fluorescent perylene derivatives, including Perylenetetracarboxylic Diimide (PTCDI), Methyl-PTCDI and Perylenetetracarboxylic Dianhydride (PTCDA). These fluorophores have been extensively studied as a material for thin film transistors and optoelectronics~\cite{dimitrakopoulos02}. The study of mono- and multilayers adsorbed on hBN has recently become a subject of intense interest with the formation of J-aggregates and superluminescence~\cite{zhao19}, excitonic coupling~\cite{kim23, juergensen23}, polarization anisotropy~\cite{kim23} and electroluminescence~\cite{zhao19,svatek20} already being reported.
\par
While some initial studies have produced perylene derivative films via solution deposition~\cite{kerfoot18}, most research has focused on growing molecular crystals of perylene derivatives using physical vapor transport~\cite{zhao19, kim23, juergensen23} or sublimation in ultra-high vacuum. These techniques offer precise control but are experimentally demanding. \par
Here we show that microspacing in-air sublimation, a recently introduced cost effective and efficient technique~\cite{He2015,ye18,guo20}, can similarly produce extended molecular crystals of PTCDI on boron nitride, while offering distinct advantages. Notably, this method provides flexibility for experimental designs requiring in-situ growth control~\cite{ye18}, all while maintaining high sample throughput. This is achieved by eliminating the need for vacuum apparatus, which not only creates space for in-situ control devices (such as microscopes) but also removes the need for evacuation cycles. \par
\section{Results and Discussion}
\subsection{Microspacing in-air Sublimation}
\begin{figure}
  \includegraphics[width = 240pt]{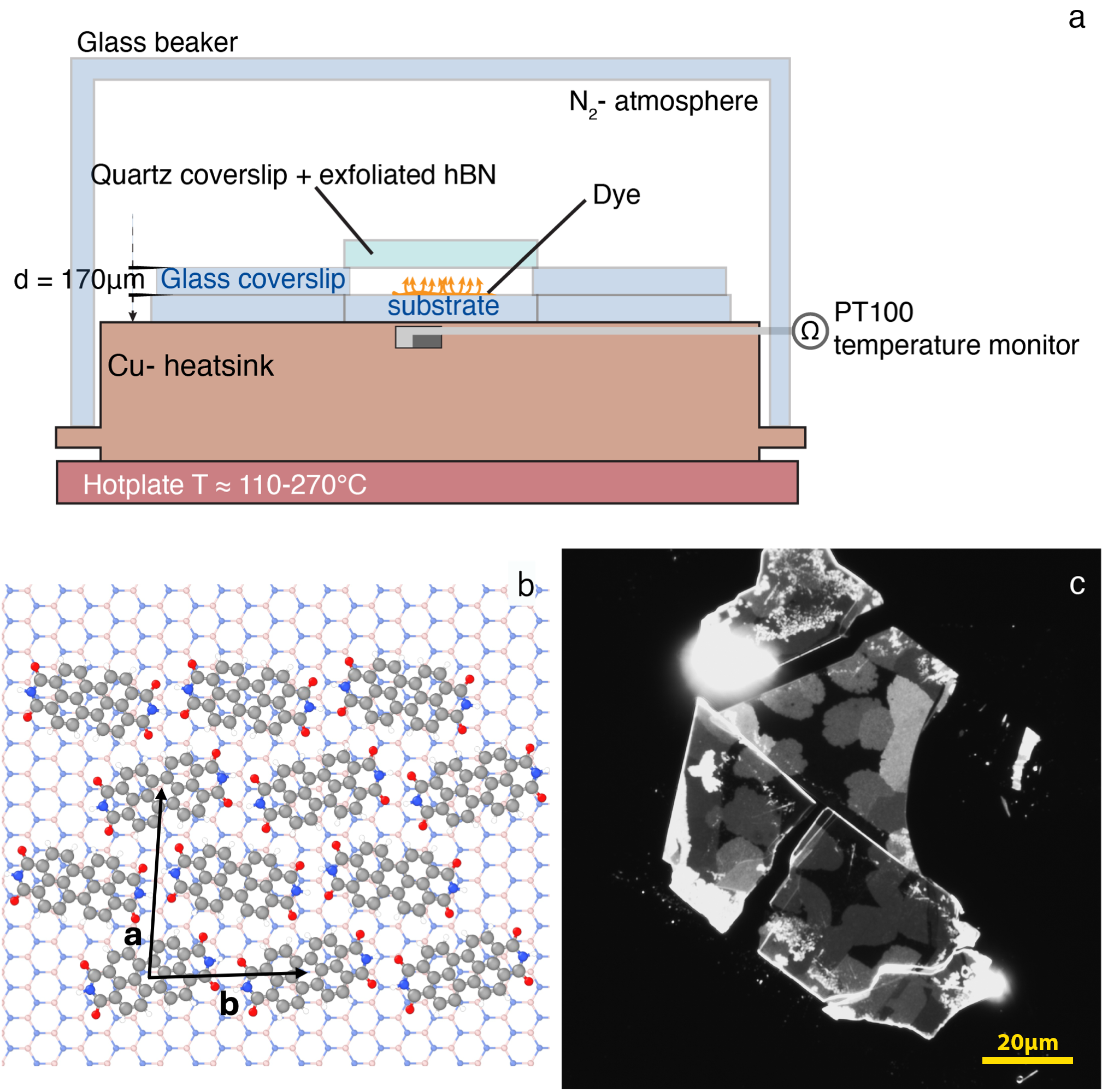}
  \caption{(a) Microspacing in-air sublimation setup. A dye-coated substrate slide is heated by a laboratory hotplate, where the heat is evenly distributed by a copper heat sink to sublimate dye molecules onto a sample coverslip on top. The air gap width is adjusted to 170~$\mu$m by utilizing standard microscopy cover glass as spacers.  (b) Molecular configuration on hBN surface revealed by Molecular Dynamics Simulations (c) Fluorescence micrograph of PTCDI evaporated onto hBN, forming extended molecular layers.}
  \label{fig:Setup}
\end{figure}
Our experimental implementation of microscpacing in-air sublimation~\cite{ye18} is presented in Fig.~\ref{fig:Setup}. We heat a  reservoir of PTCDI dye molecules, carried by a coverslip, using a standard laboratory hotplate. This reservoir is by itself a glass substrate that has been sublimation-coated with PTCDI in the same setup, using PTCDI powder as a source. This preliminary step was done to reduce inhomogeneity in the dye reservoir, further avoiding the lithographic effect described in ref. \cite{guo20}. To control the temperature with high precision and to homogenize the heat, the sublimation setup is mounted on a copper block with an internal temperature probe (PT100). The sublimation target, typically a quartz coverslip carrying freshly cleaved hBN multilayers, is mounted upside-down on the top. The gap between the reservoir and the target is adjusted to 170~$\mu$m by a coverslip as a spacer, close to prior work~\cite{ye18}. 

The entire setup is operated in a lenient nitrogen atmosphere to prevent oxidation of the dye. This is achieved by covering the sublimation setup with a glass beaker and flushing the volume within it with \(\ce{N2}\) prior to the sublimation step. While the beaker does not form a tight seal and therefore does not allow for precise control of pressure or partial water vapor levels, it effectively displaces ambient air containing oxygen. More information on the technical details of the sublimation system can be found in the Supplementary Material in Fig. S5 and Fig. S6.
Sublimating dye onto a freshly exfoliated hBN multilayer flakes by this setup produces extended molecular layers that are readily visible in fluorescence microscopy (Fig.~\ref{fig:Setup} b). Note that no polydimethylsiloxane (PDMS)- stamping is involved in the preparation of the hBN target. The hBN is directly exfoliated onto a cleaned quartz coverslip from bulk material, avoiding PDMS or glue contamination from stamping. The short interval (30–60 s) between exfoliation and sublimation ensures an exceptionally clean hBN surface for PTCDI adsorption.

\begin{figure}
    \centering
    \includegraphics[width = 240pt]{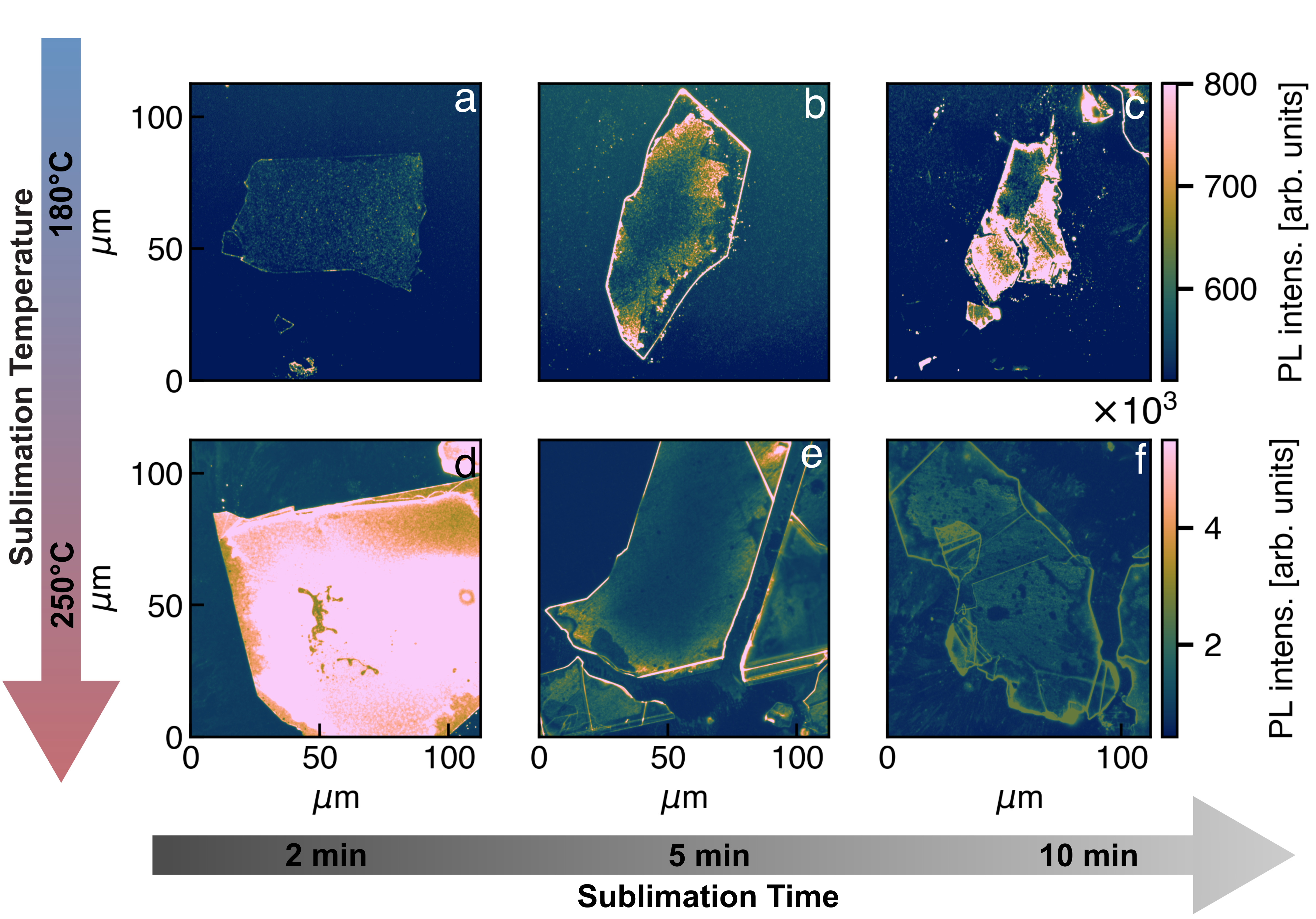}
    \caption{Fluorescence micrographs of in-air sublimated PTCDI on hBN flakes. While the samples depicted in one row were sublimated at equi-temperature. (a), (b) and (c) were sublimated at 180$^\circ$~C and (d), (e) and (f) were sublimated at 250$^\circ$~C reservoir temperature), the columns correspond to increasing exposure times of the exfoliated hBN to the molecular flux. Note the significant difference of overall fluorescence intensities between row one and two (corresponding to 180$^\circ$~C and 250$^\circ$~C sublimation temperature respectively). The units of fluorescence intensity are chosen such that the values are comparable between the two rows. With higher temperatures, the spread of overall fluorescence intensity with different sublimation times increases (see outlier in panel (d)). The gap spacing was kept constant at 170~$\mu$m.} 
    \label{fig:parameter_tailoring}
\end{figure}

To control the properties of the PTCDI layers, two experimental parameters are available: the sublimation temperature and the exposure time that the target spends in the flux of sublimating molecules.   (Fig.~\ref{fig:parameter_tailoring}~a-f). Both increasing temperature and increasing exposure time lead to increasing thickness of the molecular layers, or higher coverages in the sub-monolayer case, resulting in brighter luminescence. We occasionally observe outlying hBN- flakes within a sample with a vastly different density, like Fig.~\ref{fig:parameter_tailoring}~d, which we tentatively attribute to inhomogeneity in the dye reservoir. Note, that the surface properties of the hBN within one sample (quartz coverslip) are likely identical, as the sample preparation involves solely the exfoliation of the bulk material. Intriguingly, we note that the dye's fluorescence response tends to concentrate on the edges of the hBN flakes for long exposure time, regardless of temperature. We provisionally attribute this to heating of the hBN target, which increases mobility and diffusion of the molecules, and leads to adsorption traps at the edge. Clustering at grain boundaries, edges, and wrinkles of hBN has been observed previously~\cite{Smit2023}. Similarly, implanted fluorescent defects (color centers) in hBN tend to accumulate at these edges~\cite{Choi2016}. Our observations of molecule adsorption on hBN suggest that these edge regions play a crucial role in adsorption processes. However, the exact physical mechanisms behind this behavior remain to be fully understood.
By coarsely tuning the molecular density with exposure time and temperature, and post-selecting samples for fine control, samples with minimal coverage can be produced (Fig.~\ref{fig:parameter_tailoring}~a), which can then be screened for individual molecular layers. 

\subsection{Layered Growth of PTCDI: AFM characterisation} \label{afm_characterization}
\begin{figure}[ht]
    \centering
    \includegraphics[width = 240pt]{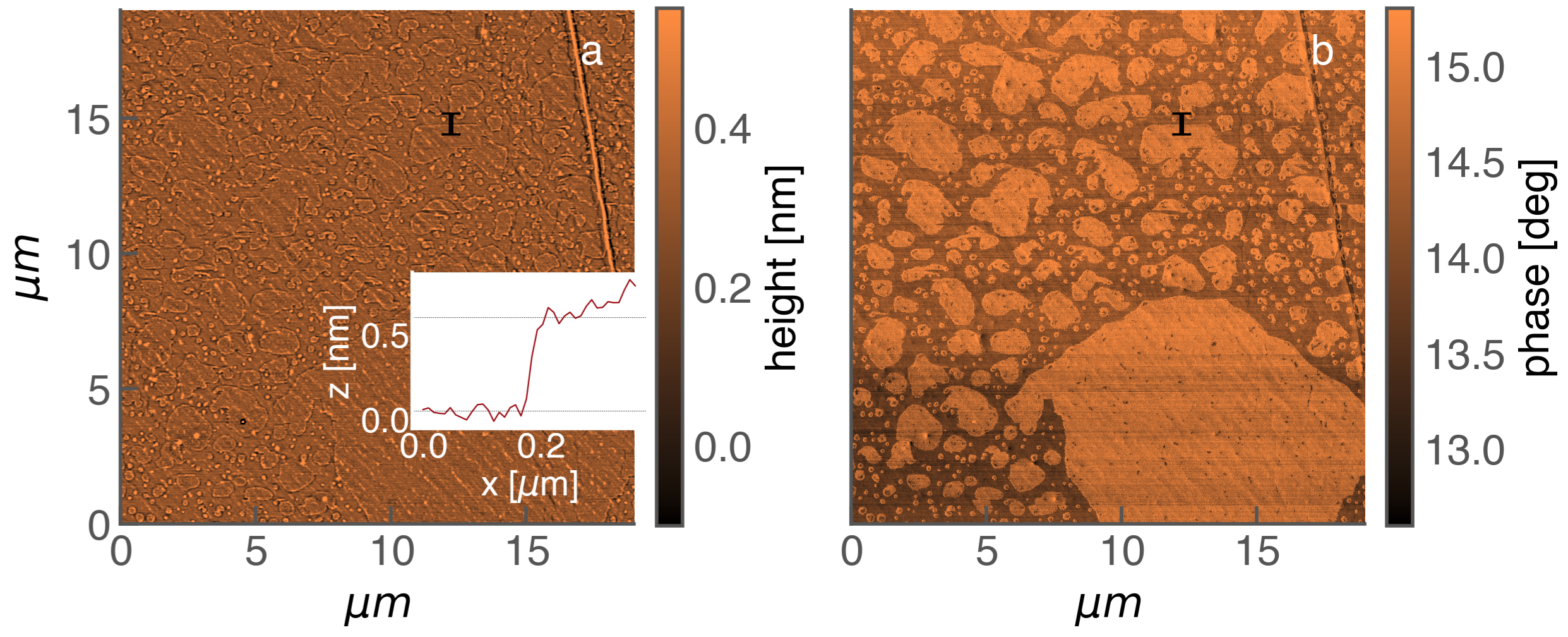}
    \caption{Tapping mode AFM scan of monolayer islands of PTCDI sublimated on freshly cleaved hBN for $t_{\text{exposure}} = 2~$min and $T=180~^\circ$C with an air gap spacing of 170~$\mu$m. (a): height image. To enhance the visibility of the island edges, the image was high-pass filtered using a difference of Gaussians. This step compensates for substrate-induced inhomogenities imprinted onto the atomically flat hBN surface, which arises from underlying substrate waviness. An unprocessed version of the AFM- height image can be found in the SI, Fig. S4. A line cut (inset), taken on the raw image at the black line in the upper right quadrant) reveals $\approx$ 0.5 nm high islands, indicative of face-on adsorbed PTCDI on hBN~\cite{Kerfoot2019}. (b): phase image. The islands of (a) appear consistently in an image the AFM feedback phase, suggesting that the islands consist of a different material (PTCDI) than the background substrate (hBN).}
    \label{fig:monolayers_AFM}
\end{figure}

One powerful technique to confirm monolayer growth is atomic force microscopy (AFM). Fig.~\ref{fig:monolayers_AFM} displays tapping-mode atomic force micrographs of a sample grown with $T=180~^\circ$C, $t_{\text{exposure}} = 2~$min. This sample has been prepared independently and is not shown in Fig.~\ref{fig:parameter_tailoring}. 
Contiguous islands are visible, which appear consistently both in the tapping-mode height and in the phase of the AFM feedback loop. This latter phase signal suggests that the material of the islands differs from the material between the islands, by having a different friction or elasticity. We interpret this as an indicator that the islands are composed of adsorbed PTCDI, whereas the area between the islands is a bare boron nitride surface. A line profile taken across an islands reveals a step height of $\approx0.5~$nm. This is slightly higher than the generally accepted value of $0.3~$nm found in similar AFM studies value for a single monolayer of face-on adsorbed PTCDI molecules~\cite{zhao19, kerfoot18_thesis}. Interestingly, this value is also inconsistent with a face-on bilayer and an edge-on monolayer, which would be expected to be at least 0.6 nm in height (see Fig.~S3). We therefore attribute the discrepancy to a difference in tip-sample interaction between the substrate and the island, which is known to make AFM height measurements unreliable \cite{novoselov04}. 

\subsection{Polarization dependent fluorescence of PTCDI multilayers} \label{PolarizationMeas}
We now turn to a more detailed study of a single sample, which features molecular domains extending over a scale of $10~\mu$m. This sample has been prepared by microspacing in air sublimation at 270$^\circ$~C for 5 min and without using a copper heatsink, a higher temperature than the previous samples, in order to achieve higher coverages of PTCDI- Islands on the hBN- substrate. This approach is substantiated by the significantly higher Photoluminescence intensity within the second - higher temperature - row in Fig. \ref{fig:parameter_tailoring} d-f. 

\begin{figure}[ht]
    \centering
    \includegraphics[width = 504pt]{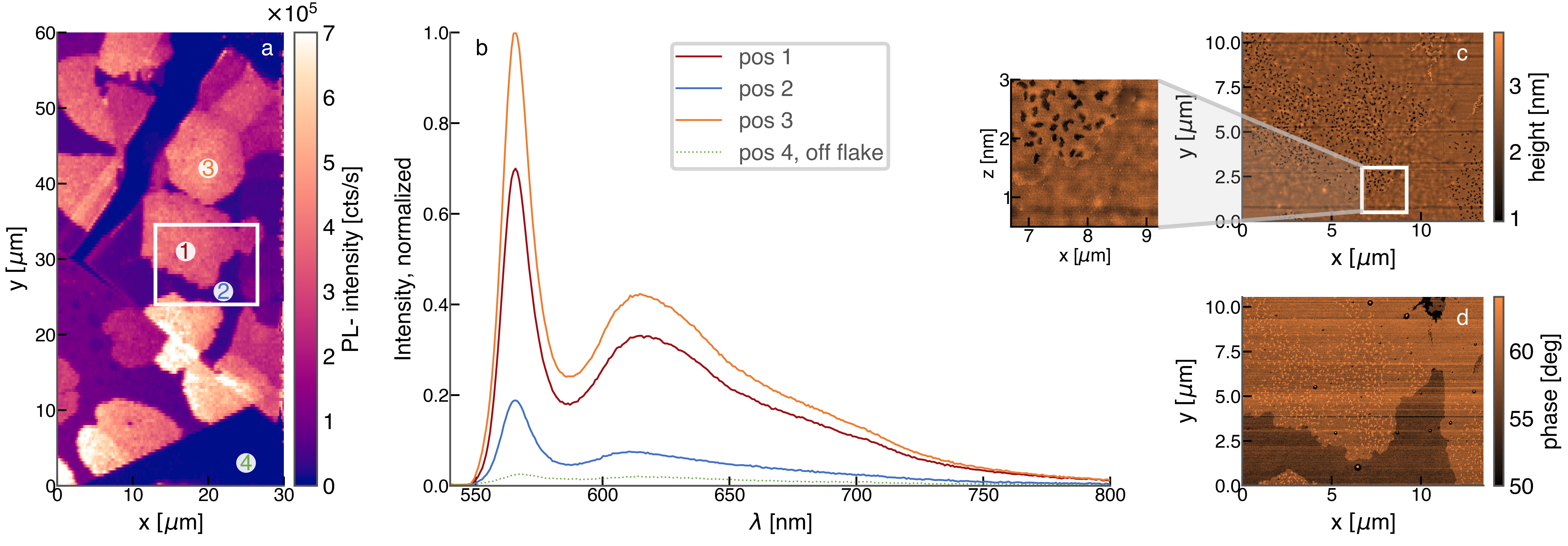}
    \caption{Correlative AFM-confocal fluorescence imaging of large PTCDI layered domains sublimated by microspacing in air sublimation at 270 °C for 5 min with an air gap spacing of 170~$\mu$m. (a) High resolution confocal fluorescence microscopy scan of a larger field of view of PTCDI on hBN. The sample region which was investigated with AFM imaging is marked with a white rectangle. Four white circles, labeled 1-4 indicate four different sampling positions of fluorescence spectrum acquisition. (b) Fluorescence spectra at four different sampling positions normalized to the maximum of fluorescence intensity at position 3. (c) High resolution AFM tapping mode scan of PTCDI on hBN, height data , median filtered for better visibility. The inset highlights a zoomed-in section of the AFM scans, showcasing holes of uniform depth within the molecular layers, located near an edge of a PTCDI- island. (d) High resolution AFM tapping mode scan of PTCDI on hBN, phase data.}
    \label{fig:AFM_fluorescence_comb}
\end{figure}

Correlative AFM-confocal photoluminescence imaging of this sample is presented in Fig.~\ref{fig:AFM_fluorescence_comb}. Molecular domains, visible as bright surfaces in photoluminescence imaging, reappear as islands of uniform height in the AFM scan. Analysis of the AFM phase confirms that the surface properties of the islands differs from the background material. A peculiar observation is the appearance of holes 
in the adsorbed islands, where the height drops by $\approx 1.5~$nm and the phase image indicates a material change with respect to both the island and the substrate. We tentatively attribute this to contamination on the flake. Due to a different phase response in these contaminated areas, the height change measured by AFM might not be reliable in these holes and the true height difference might be smaller.
Aside from this observation, all islands have the same height of approximately $0.5~$nm with respect to the background substrate, even if they differ in brightness in the photoluminescence image by up to $560~\frac{\text{kcts}}{\text{s}}$. Photoluminescence spectra taken at different spots, presented in Fig.~\ref{fig:AFM_fluorescence_comb}, appear comparable in shape to monolayer and bilayer spectra
observed for related compounds (Me-PTCDI, PTCDA) in similar work ~\cite{juergensen23,kim23,zhao19}. The spectra consistently display a narrow, intense zero-phonon line centered at 566 nm, with a full width at half maximum (FWHM) of approximately 15 nm. Accompanying the zero-phonon peak is a smaller, broader 0-1 peak, appearing at a wavelength approximately 49 nm longer. In the spectrum for ‘pos 2’ (Fig. \ref{fig:AFM_fluorescence_comb} b), the 0-1 peak is slightly blue-shifted, with a central position at 611 nm, closer to the main 0-0 peak. This spectral behavior is commonly attributed to the fluorescence of molecular mono- and bi-layers. \par

\begin{figure}[ht]
    \centering
    \includegraphics[width = 400pt]{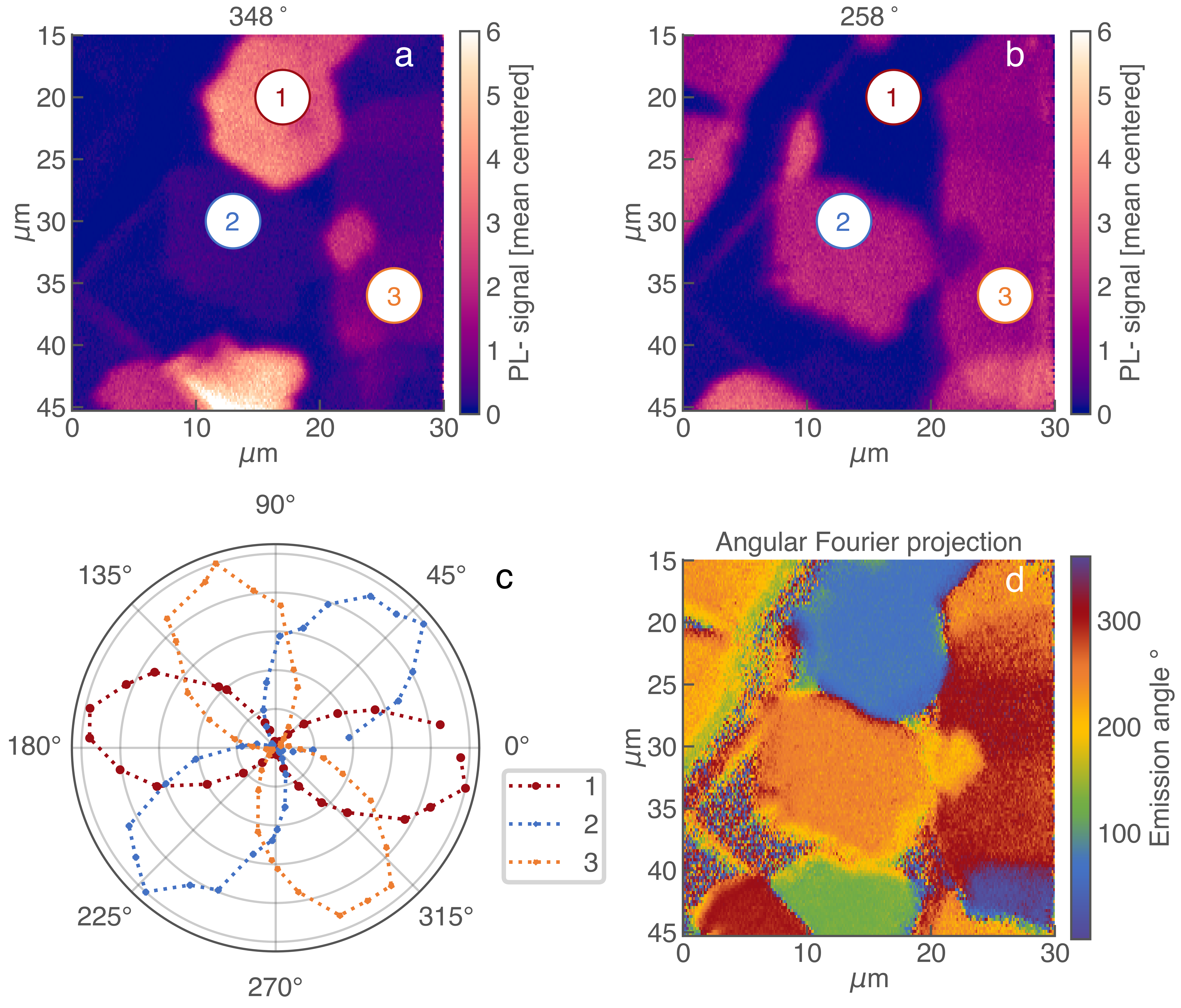}
    \caption{Polarization dependent fluorescence measurement of the PTCDI- domains presented in Fig. \ref{fig:AFM_fluorescence_comb}. The polarization was rotated 360 $\deg$ in increments of 10 $\deg$. Two representative PL images for two polarization settings are given at 348 $\deg$ (a) and 258 $\deg$ (b) respectively. Three regions of interest (ROI) (1-3) are drawn, where the average intensity within each ROI is plotted for all polarization angles (c). A map of the emission angle of each point of the imaging region (d).}
    \label{fig:parallel_config}
\end{figure}

All domains display polarized emission, confirming that the adsorbed molecules are indeed aligned with the hBN surface and long-range ordered over the entire domain. This is confirmed by  experimental results presented in Fig.~\ref{fig:parallel_config}. We excite the molecules with a linearly polarized laser at $532$~nm while we selectively detect fluorescence light of the same linear polarization. In this setting, domains differ in brightness by nearly 100\% (Fig.~\ref{fig:parallel_config} a), even those that had comparable brightness in unpolarized imaging (Fig.~\ref{fig:AFM_fluorescence_comb}). The brightness of individual domains can be altered by rotating the polarization of excitation and detection while keeping both aligned to each other. Rotating through a full circle, the domains display a dipole pattern with luminescence being maximized for excitation and detection along one particular axis and minimized for excitation and detection along an orthogonal axis (Fig.~\ref{fig:parallel_config}~c). This behavior is further visualized in Fig.~\ref{fig:parallel_config}~d, where the emission angle at each pixel is color-coded. To achieve this, a full scan across all polarization angles was performed, recording confocal images in 10° increments. A 2D Fourier transform (2D-FFT) was then applied to the resulting dataset. By extracting the phase information at the dominant angular periodicity (approximately $60^\circ$), a spatial map of the relative polarization angle at which each pixel reaches its maximum fluorescence intensity was obtained. This visualizes the 3 fold symmetry in the emission polarization angle. Different domains feature different dipole axes, rotated by multiples of $60^\circ$ against each other in a three-fold symmetric pattern, which confirms alignment of the grown PTCDI domains with the three-fold symmetric hexagonal lattice of hBN. 

\subsection{DFT Assessment of Monolayer Morphology}
Although both hBN and PTCDI feature six-membered rings, potential structures of PTCDI layers do not need to be commensurate with the underlying hBN lattice. 
Further, PTCDI  is polar and with its two imide groups acting as active moieties, it is likely to form double hydrogen bonds with neighbouring molecules. Hence the actual morphology adopted by the PTCDI monolayer is not obvious. To shed light on this issue we have performed density functional theory (DFT) based simulations of the adsorption and formation energetics. Starting point are five structures, which have been suggested for the adsorption of PTCDI on different substrates including graphene \cite{ptcdi_graphene_2014}, NaCl(001) \cite{ptcdi_NaCl_2009} and Au(111) \cite{mura2009,silly2008_gold}. To find energetically favourable structures, formation energy calculations were conducted for molecular monolayers in  vacuum, i.e. excluding the hBN layer. While the motivation comes from the much reduced computational effort, this approximation is rather reasonable as well. First, hBN is known to be a weakly interacting substrate with little influence on the electronic structure of the adsorbate \cite{melani2022}. Second, the aggregation of PTCDI is mainly driven by interactions between molecules due to the hydrogen bonding.

\begin{figure}
    \centering
    \includegraphics[width = 400pt]{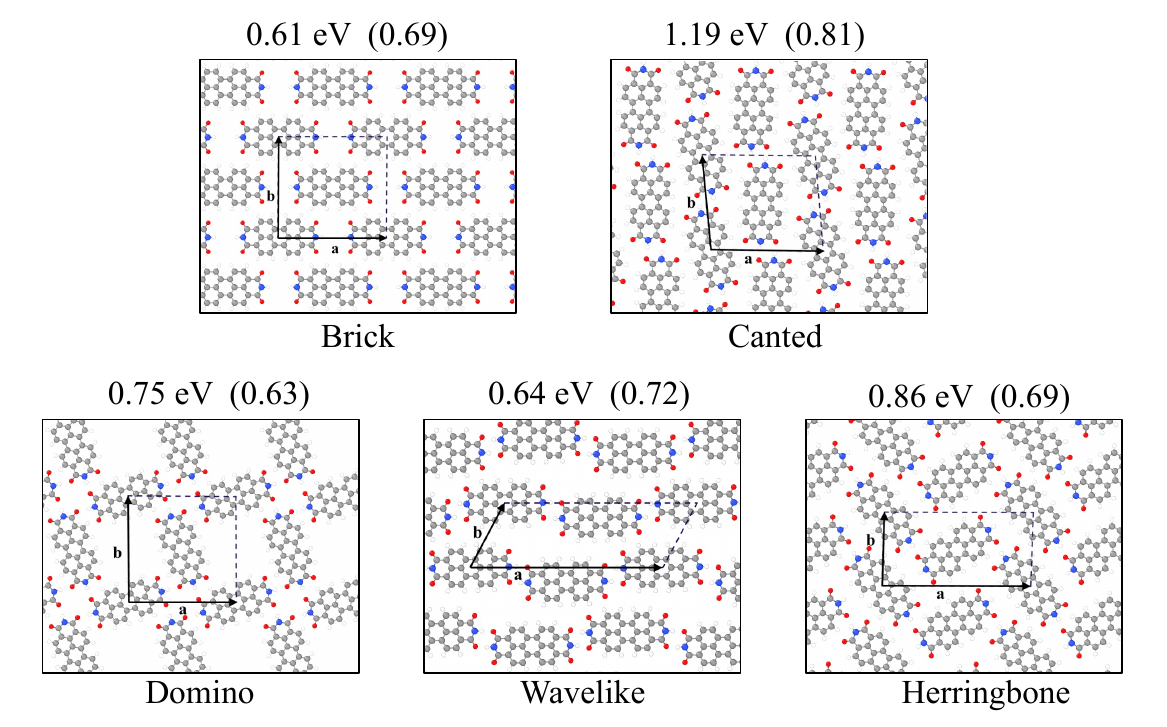}
    \caption{Comparison of five different molecular monolayers suggested on hBN, each depicted with their calculated formation energies in eV/molecule, and packing density values in parentheses above each structure. The 2D lattice vectors ${\bf a}$ and ${\bf b}$ are given for each structure.} 
    \label{fig:monolayer_models}
\end{figure}

Comparison between different cases in Fig. \ref{fig:monolayer_models} shows that the canted structure is the most stable one, followed by the wavelike structure. Since the structures are all planar and thus $\pi-\pi$ stacking doesn't play a role, differences in stability arise essentially from hydrogen-bonding.
Tighter hydrogen-bonding is enabled by the denser packing of the canted structure as seen from the packing densities reported in Fig. \ref{fig:monolayer_models}.

\begin{figure}
    \centering
    \includegraphics[width = 400pt]{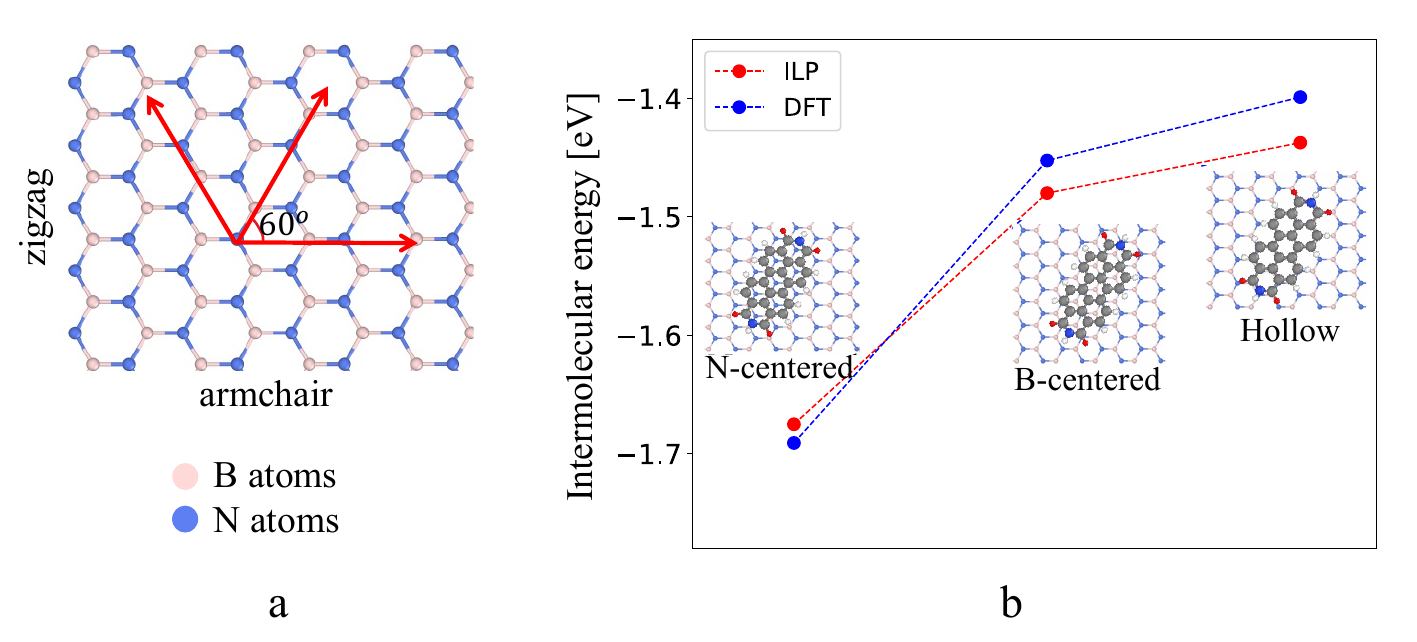}
    \caption{(a) View of hBN structure, showing its zigzag and armchair directions, with the three arrows indicating distinct armchair directions. (b)  Energetics of potential  adsorption sites for a single PTCDI molecule on the hBN surface at equilibrium distance as calculated by DFT and ILP for comparison.} 
    \label{fig:dir_dep_PTCDI}
\end{figure}

To further analyze adsorption geometries, we identified three adsorption sites for a single PTCDI molecule on the hBN as shown in Fig. \ref{fig:dir_dep_PTCDI}. The results evidence that the molecules preferentially occupy the N-centered configuration along one of the armchair directions of hBN, aligning their long molecular axis accordingly. This observation is in line with the polarization dependent fluorescence data shown in Fig. \ref{fig:parallel_config}. 
Figure \ref{fig:dir_dep_PTCDI} contains two other configurations, i.e. the B-centered and hollow one. They are energetically less favoured by 
0.23~eV and 0.29~eV, respectively. Note that while the equilibrium distance for the N-centered case is 0.340~nm, it increases to 0.342~nm and 0.346~nm for the  B-centered and hollow case, respectively.
We have also inspected the lateral potential energy surface resulting from shifting the molecule in both armchair and zigzag directions. The energetics of this  surface also depends on the distance. At the N-centered equilibrium distance the highest and lowest lateral corrugation of the surface (energy barrier) is 0.26~eV and 0.19~eV, respectively. 
However, these barrier reduce to 0.18~eV (30 \% reduction) and 0.15~eV (27 \% reduction) if the molecule-surface distance at the barrier position is energetically optimized.
Despite such high energy barriers compared to thermal energy at room temperature, a single molecule remains mobile on the surface due to thermally activated processed, causing a hopping between high adsorption sites \cite{dobbs_dynamics_1992}. This in turn enables formation of aggregates and eventually monolayer crystals as seen in the experiment.

To shed light into the aggregation process we have performed Molecular Dynamics (MD) simulations of medium-covered (66 $\%$) hBN surfaces using a recently developed interlayer potential (ILP, Tab.~S1); details can be found in the Supplementary Information. We note that this ILP gives reasonable agreement with DFT for the adsorption site energies, see Fig. \ref{fig:dir_dep_PTCDI}. For reference, the case of a single PTCDI molecule on hBN has been considered as well. In the Supplementary Material, Fig. S1, we show the trajectory data illustrating  the relation between the molecule's vertical distance above the surface and the adsorption energy. This figure illustrates the importance of vertical thermal motion, reducing the adsorption energy and thereby enabling lateral motion. The obtained 2D diffusion constant is $(8.6 \pm 0.1)$ $\times$ 10$^{-9}$ m$^2$/s.

\begin{figure}
    \centering
    \includegraphics[width = 460pt]{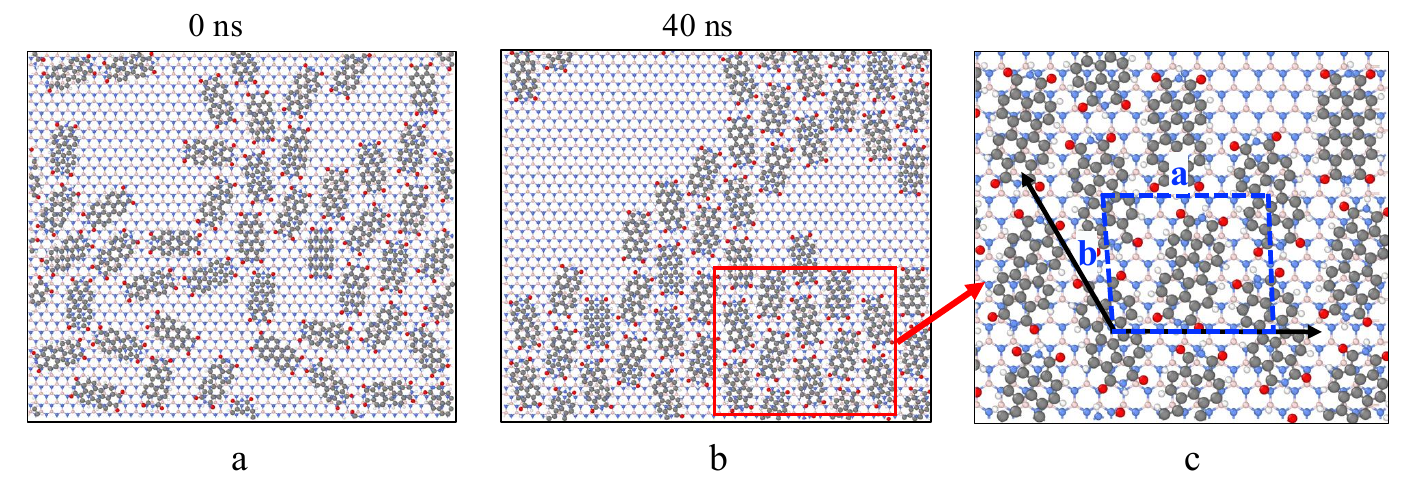}
    \caption{ Molecular dynamics snapshots of PTCDI adsorption on the hBN surface, illustrating the structural evolution over time. (a) Initial configuration at 0 ns, showing dispersed PTCDI molecules on the surface; (b) final configuration at 40 ns, where PTCDI molecules have formed a stable aggregated layer; (c) enlarged view of the crystalline structure, highlighting molecular alignment and lattice parameters. The black arrows show the lattice vectors of hBN.}
    \label{fig:PTCDI_morphology}
\end{figure}

Snapshots of two structures obtained from MD simulation are presented in Fig \ref{fig:PTCDI_morphology}. Initially, PTCDI molecules were randomly deposited from a distance of 3~nm onto the hBN monolayer. After the desired surface coverage was reached (0 ns) an annealing process was carried out at 600 K for 40 ns to accelerate nucleation. As expected from the DFT calculations, the molecules formed a canted arrangement, with unit cell parameters of  $a=1.43$ nm, $b=1.65$ nm and $\alpha=89^\circ$. The canted arrangement is stable even with respect to thermal fluctuations. The analysis further reveals orientation locking of the PTCDI molecules into a specific orientation with respect to the hBN lattice (e.g. vertical alignment in Fig.~\ref{fig:PTCDI_morphology} b), consistent with the experimental observation of orientation-locking to the lattice. We also observe coincidence epitaxy of the PTCDI, in the sense that molecules spaced by small integer multiples of lattice vectors of the molecular lattice adsorb on the same (e.g. N-centered) kind of site of hBN. A more detailed study of the epitaxy is presented in the Supplementary Information (Fig. S2, Table S2).

\section{Conclusions}
In summary, we have demonstrated that a facile technique, microspacing in-air sublimation, can grow controlled layers of monocrystalline aligned organic dyes on a boron nitride substrate. The resulting layers display comparable spectral and polarization anisotropy properties to samples prepared by more sophisticated schemes like physical vapor transport~\cite{juergensen23,kim23}. At the same time, they display extended molecular islands of tens of microns size, which display highly polarized fluorescence emission along three principal emission axes. In order to scrutinize possible morphologies of PTCDI on hBN, DFT and force field MD simulation have been performed, which pointed to the preferential growth of canted-like structures. 

We anticipate microspacing in-air sublimation to become a versatile platform for studies of collective emission in molecular layers~\cite{juergensen23}, electroluminescence~\cite{svatek20}, and quantum technology studies using organic fluorophores as optically readable qubits~\cite{neumann23, mena24}.  

\section{Experimental Section}
\subsection{Tapping Mode AFM- scans}
All AFM- scans exhibited in this study are done using a \textit{Park Systems, NX20} AFM combined with premounted AC160TS- cantilevers ($f_{\text{resonance}} \approx 300~$kHz, $r_{\text{tip}} \approx 7~$nm and $k\approx 26$ N/m). Even though, operating an AFM in tapping mode is an interacting, sometimes destructive imaging mode, the surface- tip interaction reveals valuable surface property changes over the scanned areas, enabling e.g. the detection of material changes by evaluating the phase image of a scan (see section\textit{ Layered Growth of PTCDI: AFM characterisation} \ref{afm_characterization} and Fig. \ref{fig:monolayers_AFM}). Due to the experimental setup of this multi-method study, the samples were mounted onto the AFM’s sample holder using vacuum suction. This approach, however, 
couples in vibrations produced by the vacuum pump, introducing mechanical noise. This ultimately limits the ability to achieve higher resolution scans, such as room temperature AFM- measurements resolving molecular configuration \cite{kerfoot18}.
\subsection{Fluorescence Imaging: Confocal and widefield microscopy}
Confocal fluorescence measurements were performed using a high-sensitivity, custom-built confocal fluorescence microscope, designed to image single fluorescent emitters with single-photon detection capability via an avalanche photodiode (Excelitas SPCM-AQRH). Scanning was achieved by rastering the back focal plane of the microscope objective (Olympus UPlanApo 60x/1.42 Oil, FN 26.5) with an angled tip-tilt mirror. \par
Widefield fluorescence micrographs are captured using a commercially available fluorescence microscope (Nikon Eclipse LV100ND). The temperature-time series of Fig.~{\ref{fig:parameter_tailoring}} has been obtained by evaporating onto six different exfoliated hBN samples and picking a reasonably large flake on each as a representative spot for the presented widefield micrographs. 


\subsection{Computational Details}

All DFT calculations have been performed using the Vienna Ab initio Simulation Package (VASP) \cite{kresse1993,kresse1994,kresse_efficiency_1996,kresse1996} with periodic boundary conditions. The exchange-correlation interactions were treated using the Generalized Gradient Approximation (GGA) in the Perdew-Burke-Ernzerhof (PBE) functional\cite{pbe_dft}. Projector Augmented Wave method with a plane-wave energy cutoff of 400 eV was used. To account for long-range van der Waals (VdW) interactions, the Many-Body Dispersion correction\cite{tkatchenko2009} was applied. 
Classical MD simulations were performed using GAFF\cite{wang_gaff_2004} force field for intramolecular interactions of PTCDI molecules. For the intermolecular interactions, we applied our recent parametrization of an Interlayer Potential (ILP) following the idea of ref. \cite{leven_2014}, designed for interactions between hBN and planar organic molecules. All simulations were carried out using the LAMMPS code \cite{plimpton_fast_1995}, in which the ILP package is implemented \cite{ouyang_2018}.

\begin{acknowledgement}
This work has been supported by the Deutsche Forschungsgemeinschaft (DFG, grants RE3606/1-2, excellence cluster MCQST EXC-2111-390814868 and SFB 1477 “Light–Matter Interactions at Interfaces” (Project No. 441234705). K.W. and T.T. acknowledge support from the JSPS KAKENHI (Grant Numbers 21H05233 and 23H02052) , the CREST (JPMJCR24A5), JST and World Premier International Research Center Initiative (WPI), MEXT, Japan. The authors thank Julian Schröer and Lisa Böhme for helpful discussions and experimental training. 
\end{acknowledgement}

\begin{suppinfo}
Supporting information includes details of the ILP parametrization as well as further trajectory data describing PTCDI diffusion on hBN.

\end{suppinfo}

\bibliography{literature}

\begin{figure}
    \centering
    \includegraphics[width = 3.25in]{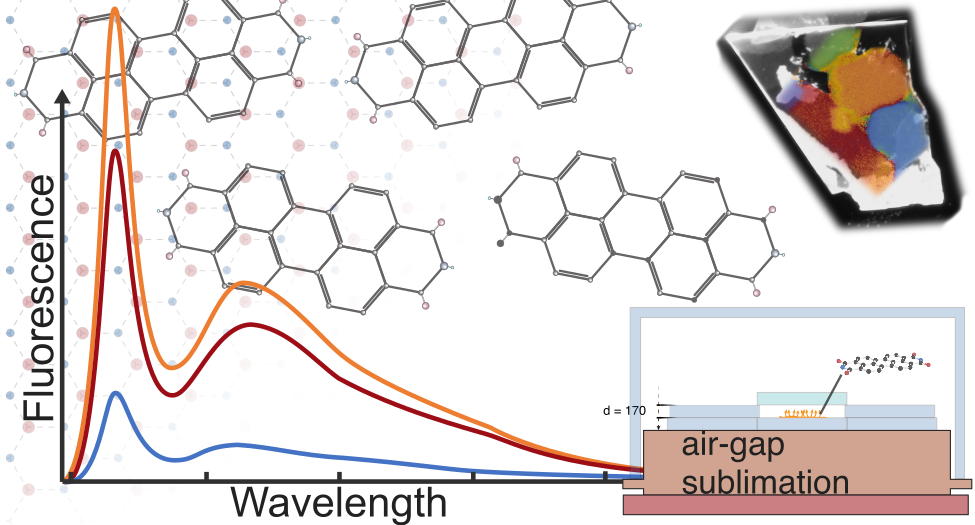}
    \caption{Table of Contents}
    \label{fig:TOC-Graphic}
\end{figure}
\end{document}